\documentclass[aps,prl,twocolumn,showpacs,superscriptaddress,floatfix,amssymb,amsmath,showpacs]{revtex4}
\usepackage{graphicx}
\usepackage{dcolumn}
\usepackage{bm}
\usepackage{color}
\newcommand{\be}{\begin{equation}}
\newcommand{\ee}{\end{equation}}
\newcommand{\bea}{\begin{eqnarray}}
\newcommand{\eea}{\end{eqnarray}}

\newcommand{\vare}{\varepsilon}

\newcommand{\re}{\mbox{e}}
\newcommand{\ba}{\begin{array}}
\newcommand{\ea}{\end{array}}
\def\nn{\nonumber\\}
\def\vare{{\varepsilon}}
\def\up{\uparrow}
\def\down{\downarrow}

\begin{document}
  
  \title{Enhancement of the Kondo effect through Rashba spin-orbit interactions} 
  
 \author{Mahdi Zarea}
   \affiliation{Department of Chemistry, Northwestern University, Evanston, Illinois 60208 }
   \affiliation{Department of Physics and Astronomy, Nanoscale  and Quantum Phenomena Institute, and Condensed Matter and Surface Science Program,\\Ohio University, Athens, Ohio 45701}
\author{Sergio E. Ulloa} 
\author{Nancy Sandler}
  \affiliation{Department of Physics and Astronomy, Nanoscale  and Quantum Phenomena Institute, and Condensed Matter and Surface Science Program,\\Ohio University, Athens, Ohio 45701}

 \date{\today}

 \begin{abstract}
We study a one-orbital Anderson impurity in a two-dimensional electron bath with Rashba spin-orbit interactions in the Kondo regime. The spin SU(2) symmetry breaking term couples the impurity to a two-band electron gas. A Schrieffer-Wolff transformation shows the existence of the Dzyaloshinsky-Moriya (DM) interaction away from the particle-hole symmetric impurity state. A renormalization group analysis reveals a two-channel Kondo model with ferro- and anti-ferromagnetic couplings. The parity breaking DM term renormalizes the antiferromagnetic Kondo coupling with an exponential enhancement of the Kondo temperature.
\end{abstract}

\maketitle

The search for new materials and architectures appropriate for spintronic and quantum computing devices \cite{Wolf} 
has renewed the interest on the study of Rashba spin-orbit interactions (RSO) \cite{Rashba1} in low-dimensional systems. For systems on surfaces, the natural lack of inversion symmetry makes RSO ubiquitous, and particularly relevant for studies of surface-related magnetic properties \cite{Rashba2}. Research on this area has been driven by rapid developments of STM techniques that have made possible the design and manipulation of atomic structures on surfaces, and study charge and spin physics in two-dimensional (2d) systems.  Among the many structures of interest, several groups have focused on the physics of isolated magnetic impurities on metallic substrates to investigate signatures of  Kondo physics in two dimensions \cite{Crommie1,Manoharan1,Manoharan2,Crommie2, Crommie3,Kern1,Crommie4,Saw1,Kern3,Crommie5,Saw2}. While these studies have unveiled new physics, they have not addressed the relevance of spin-orbit (SO) interactions possible on many metallic substrates serving as reservoirs for the magnetic impurity. The fact that these interactions can profoundly modify the spin structure on a surface was demonstrated in spin-polarized measurements of Mn impurities on W substrates \cite{Bode}. Images and characterization of beautiful chiral spin ordered structures are indeed understood in terms of the strong SO coupling on the surface. Moreover, recent investigations of magnetic impurities on graphene \cite{Manoharan3,Crommie6}, the ultimate two-dimensional system, pose the question of the influence of the underlying graphene substrate on measured magnetic properties.

The role of SO interactions in the Kondo regime of magnetic impurities embedded in metallic hosts has been a topic of debate since the early measurements of electrical resistivity in Cu:Mn compounds doped with Pt, carried out by Gainon and Heeger \cite{Heeger}. The reduced divergence in resistivity, interpreted as due to the presence of SO scatterers, was taken as evidence for the suppression of the Kondo effect. Much theoretical and experimental activity followed with rather inconclusive results \cite{Giovan-Everts-Bergmann-Wei-Vangrun-VanHaese-Akiyama}: while some works supported Gainon and Heeger's findings, others reached opposite conclusions. Noting that SO interactions preserve time-reversal symmetry, Meir and Wingreen \cite{Meir} showed that the Kondo regime is unaffected by SO interactions in the infinite Hubbard-U limit. More recently, however, a solution for a two-dimensional Kondo model in the presence of SO interactions predicts that the Kondo temperature increases by a multiplicative factor proportional to the SO coupling constant \cite{Malecki}. Thus, the question: {\it what is the role of SO interactions in the Kondo regime for two dimensional systems?} remains controversial. The purpose of this Letter is to provide a definitive answer by presenting a solution to the model of a one-orbital Anderson magnetic impurity \cite{Hewson} embedded in a two-dimensional metallic host with RSO interactions. We show that the presence of these interactions reduces the Anderson Hamiltonian to an effective two-band Anderson model coupled to the impurity. By an appropriate Schrieffer-Wolff transformation \cite{SW}, the Hamiltonian reduces to an effective two-channel Kondo model plus a Dzyaloshinsky-Moriya (DM) \cite{DM} interaction term. The renormalization group analysis for this effective Hamiltonian reveals that the impurity couples to the bath with ferro- and antiferromagnetic couplings, with the one-channel Kondo model as the fixed point at low energies. More surprisingly, as we show below, the presence of DM interactions, which vanish at half-filling and at the Hubbard $U$-infinity limits, introduces an {\it exponential} increase in the value of the Kondo temperature.

{\it The model.} 
The Anderson Hamiltonian of a two dimensional electron gas in the
presence of RSO interactions is given by $H = H_0 +  H_U+ H_{hyb} + H_{RSO}$. Here
 $ H_0= \sum_{\bm k s}\vare_{k}c_{\bm k s}^{\dag}c_{\bm k s} 
+\sum_s\vare_{d}c_{d s}^{\dag}c_{d s}$ describes free electrons with momentum ${\bm k}$ and spin $s$, and one spherically-symmetric 
impurity level with energy $\vare_{d}$. The Hubbard interaction term  is given by $H_U =U n_{d\up}n_{d\down}$ with
$n_{ds}=c_{ds}^{\dag}c_{ds}$ as the  impurity electron density. 
The hybridization term $H_{hyb} = \sum_{\bm k s} V_{\bm k} c_{\bm k s}^{\dag} c_{d s} + h.c.$ describes  the coupling between the 
 impurity and bath electrons in the absence of the RSO interaction.  We consider spin $1/2$ electrons for the bath and the impurity site and (without loss of generality) $V_{\bm k}$ to be real and independent of the electron spin. The RSO interaction is described by:
\bea
\label{RSO}
H_{RSO}&=&\sum_{\bm k}\lambda_{R} (k_y+ik_x )c_{\bm k\up}^{\dag}c_{\bm k\down}+h.c. \nonumber \\
&=&\sum_{\bm k}\lambda_{R} k\re^{-i\theta_{\bm k}} c_{\bm k\up}^{\dag}c_{\bm k\down}+h.c.
\eea
with $\theta_k$ defined by $k_x=-k\sin\theta_{k}$,~$k_y=k\cos\theta_{k}$, and $k=|\vec k|$. This model allows one to study the effect of the RSO with variable strength  $\lambda_{R}$.
Notice that the effect of {\em local} RSO interactions (acting only at the impurity atom) has been the topic of previous studies \cite{Sun,Simonin,Lopez,Ding,Vernek,Lim,Lucignano} but it is not considered in the present model. An experimental setup where this model is realized consists for example of isolated magnetic impurities (Co, Mn, etc.) deposited on the surface of materials with high RSO interactions such as Au(111). 

In the angular momentum basis electron operators are:
\be
\label{eq:m_base}
c_{{\bm k}s}=\sum_{m = -\infty}^{m= \infty}\frac{\re^{im\theta_{k}}}{\sqrt{2\pi k}}c_{ks}^m \;\;;\;
c_{ks}^m=\sqrt{k\over2\pi}\int d\theta_{k} \re^{-im\theta_{k}}c_{{\bm k}s} . 
\ee

The canonical transformation:
\bea
&&c^{m+1/2}_{k h}=\left(c^{m}_{k \up}
+ h c^{m+1}_{k \down}\right)/\sqrt{2}
\label{eq:bath-ops}
\eea
diagonalizes the bath Hamiltonian in the presence of RSO.\@ Here $h = \pm 1$ and $j=m+h/2$  are the chirality and angular momentum quantum numbers, respectively. These operators satisfy standard anticommutation relations:
$\{(c^{m+1/2}_{k h})^{\dagger}, c^{m'+1/2}_{k' h'}\} = \delta_{k,k'} \delta_{h,h'} \delta_{m+1/2; m'+1/2}$.

The  corresponding band energies are $\vare_{k h}=\vare_k+\lambda_R k h$. After the transformation, the total Hamiltonian is:
\bea
H =\sum_{k h m} \vare_{k h} \left(c_{k h}^{m+1/2}\right)^{\dagger} c_{k h}^{m+1/2} + H_{imp}+ \sum_{k h m} \frac{\tilde{V}_{k} \delta_{m, 0}}{\sqrt{2}}\times &&
\nonumber \\
\left [ \left (c_{k h}^{m+1/2}\right)^{\dagger} c_{d \up} + (-1)^{(\frac{1-h}{2})}\left(c_{k h}^{m-1/2}\right )^{\dagger} c_{d \down}+h. c.\right] &&
\label{eq:total H}
\eea
where $H_{imp}=\sum_{s} \vare_{d} c_{d s}^{\dagger} c_{d s} + H_{U}$;  $c_{ds}$ is the operator for the local orbital in the angular momentum basis and $\tilde{V}_{k} = V_{k}\sqrt{\frac{2\pi}{k}}$.
Thus, the  RSO term produces an effective two-band ($h=\pm 1$) Anderson problem with the impurity coupled to $j=\pm 1/2$ channels in each band. 

{\it The Kondo regime.}
To describe the Kondo regime we perform a Schrieffer-Wolff transformation (SWT) \cite{SW}. As in the usual one-impurity Anderson model, the SWT is obtained by requiring  the effective Hamiltonian $H_{{\rm eff}}=\re^S H \re^{-S}$ not to contain an $H_{hyb}$ term to first order. The resulting $H_{\rm eff}$ contains $H_{0}+H_{U}$ with renormalized parameters, plus an exchange Hamiltonian. Using Eqs.\  \ref{eq:bath-ops} and \ref{eq:total H}, we find $S$ to be given by \cite{Supplementary}:
\bea
\label{eq:S}
&&S=\sum_{k h} T_{k h}\left[ \left(c_{k h}^{1/2}\right)^{\dag} c_{d \uparrow} + \left(c_{k h}^{-1/2}\right)^{\dag} c_{d \downarrow} \right] - h.c. 
\nn&&T_{k h} = V_{k}(  n_{d \bar h} G_{k h}+  g_{k h})
\nn&&G_{k h} = {1\over \vare_{k h}-\vare_{d}-U} 
-{ 1\over \vare_{k h}-\vare_{d}} 
\nn&&g_{kh}={ 1\over \vare_{k h}-\vare_{d}}\, ,
\eea
where  we defined $n_{d\bar h} = n_{d\downarrow}  (n_{d\uparrow})$ for $h=1 (h=-1)$, and $\bar h = -h$. As in the usual case, $S$ involves the Green's functions of free particles moving in a bath that contains the localized impurity level.
Notice that this transformation reduces to the standard form for the SWT, written in the chiral basis, when the RSO coupling is zero ($\lambda_{R}=0$).   
The SWT reveals that the bath fermions relevant in the Kondo
regime are:
\bea
c_{k \pm}^{1/2} &=& \left(c_{k \up}^{0} \pm c_{k \down}^{1}\right)/\sqrt{2} \nonumber \\
c_{k \pm}^{-1/2} &=& \left(c_{k \up}^{-1} \pm c_{k \down}^{0}\right)/\sqrt{2} 
\label{eq:bathfermions}
\eea
emphasizing the conservation of the total angular momentum in the $z$-direction: a spin flip process is
compensated by changes in the orbital angular momentum channels $m=0, \pm 1$.  Once the relevant modes are identified, it is
more convenient to return to the original basis to describe band-electrons and introduce standard spinor notation. 
The resulting effective  Hamiltonian contains two different terms:
$H_{\rm eff}=\sum_{\bm {k, k'}}({\cal H}_{K}+{\cal H}_{DM})$. The first, equivalent to the standard Kondo Hamiltonian is:
\be
\label{eq:H_K}
{\cal H}_{K}= J_{k k'} (\bm s_{\bm {k k'}} \cdot \bm S 
- \frac{1}{4}\rho^c_{k k'} \rho^d ),
\ee 
where
$ \bm s_{{\bm {k k'}}}= {1 \over 2}c^{\dag}_{\bm k s}\bm \tau^{s s'} c_{\bm k' s'}$,
$ \rho^c_{{\bm {k k'}}}=  c^{\dag}_{\bm k s}\tau_0^{ss'}c_{\bm k' s'}$. The corresponding definitions for the
impurity operators are: $ \bm S = {1\over2}c_{d s}^{\dag} \bm \tau^{ s s'} c_{d s'}$ and $\rho^d = c_{d s}^{\dag} \tau_0^{s s'}c_{d s'}$.
Here $(\bm \tau,\tau_0)$ are the standard Pauli matrices. 
The Kondo coupling is given by 
$J_{k k'}= -V_{k}V_{k'} (G_{k +} + G_{k -} + G_{k' +} + G_{k' -})/2$, 
it is averaged over the chirality quantum number  $h$ and is spin-independent. One can verify that ${\cal H}_{K}$ reduces to the standard Kondo Hamiltonian when $\lambda_{R}=0$. In the angular momentum basis (\ref{eq:m_base}) this Hamiltonian reads
${\cal H}_{K} \sim J_{k k'}  \sqrt{kk'} ( \bm s_{k k'}^{0} \cdot \bm S 
-{1\over4} \rho^c_{k k'}  \rho^d )$ with $\bm s_{k k'}^{0}={1\over2} c^{0\dag}_{ k s}\bm \tau^{s s'}c^{0}_{k' s'}$
and $ \rho_{k k'}=c^{0\dag}_{k s} \tau_0^{s s'}c^{0}_{k' s'}$, i.e., involving only the $m=0$ mode. In this expression the angular integration has already been carried out.
 
After some algebra \cite{Supplementary}, the expression for the second term, ${\cal H}_{DM}$, is:
\be
\label{eq:DM}
{\cal H}_{DM} =
i \lambda_R k_{F}C({\bm k}-{\bm k'})\cdot (\bm{s}_{{\bm {k k'}}} \times \bm{S})
\ee
that corresponds to the Dzyaloshinsky-Moriya (DM) interaction \cite{DM}. Here $ \bm{k} -\bm{k'}$ is a vector in the 2d plane, $\bm{s_{\bm {k k'}}}$ and $\bm{S}$ are the bath and impurity spin vectors respectively. The expression has been evaluated for scattering processes near the Fermi surface where $\vare_{kh} \approx 0$ and $k \approx k' = k_{F}$ up to first order in $\lambda_{R}$, with $C=C(\vare_{d}, U)$. It is important to remark that  $C$ vanishes in the limit of $U \to \infty$ and when the impurity state is particle-hole symmetric, i.e. $\vare_d = -U/2 $, in agreement with general time-reversal symmetry arguments \cite{Meir, Supplementary}. 
Similar terms were found in previous works \cite{Xia}, 
when the hybridization coupling $V_{k}$ is made to be spin-dependent, i.e, $V_{k s}$. However in these models
the DM term does not vanish at the particle-hole symmetric point as expected.  

It is also instructive to write the DM term in the angular momentum basis (using Eq.\ \ref{eq:m_base} and carrying out the angular integration part), where it reads:
${\cal H}_{DM}=  \lambda_{R} k_F C (\bm s^{\lambda}_{k k'} \cdot {\bm
  S}-{1\over4}\rho^{\lambda}_{k k'}\rho_d)$, 
where $\bm s^{\lambda}_{k k'}={1\over2}(c^{0\dag}_{ k s}\bm \tau^{ss'}c^{2s'}_{k' -s'}
+c^{2s\dag}_{ k-s}\bm \tau^{ss'}c^{0}_{ k' s'} )$, $ \rho^{\lambda}_{k k'}=(c^{0\dag}_{k s} \tau_0^{ss'}c^{2s'}_{k' -s'}
+c^{2s\dag}_{k -s}\tau_0^{ss'}c^{0}_{k' s'})$.
It is clear that this interaction couples the $m=0$ and $m=\pm1$ modes of band electrons (as $c^{2s}_{k-s}=c^{1}_{k\downarrow}$ for $s={1\over2}$, etc.).

{\it Renormalization group analysis.} 
To understand the effect of the DM terms in the Kondo regime we perform a renormalization group (RG) analysis, assuming that $V_{k} \approx V_{k_{F}}$ and $J_{k k'}  \approx J_{k_{F}}$. The RG flow reveals the existence of an additional term involving higher energy bands: 
$H_{\gamma} = \gamma (\bm s^{\gamma}_{\bm {k k'}}. \bm S - {1\over4}\rho^{\gamma}_{\bm {k k'}} \rho_d)$ introducing a new coupling constant $\gamma$. 
In the angular momentum basis (after angular integration), the expression for $\bm s^{\gamma}_{\bm{k k'}}$ reads: $\bm s^{\gamma}_{kk'}=c^{2s\dag}_{ k-s}\bm \tau^{ss'}c^{2s'}_{k'-s'}$ and  $ \rho^{\gamma}_{kk'}=c^{2s\dag}_{k-s} \tau_0^{ss'}c^{2s'}_{k'-s'}$. The resulting coupled equations are:
\bea
\dot J&=&J^2+\lambda_F^2/4
\nn \dot \gamma &=& \gamma^2+\lambda_F^2/4
\nn \dot \lambda_F &=&(J+\gamma)\lambda_F \, ,
\eea
where  $\lambda_F = 2 C \lambda_{R} k_F$. These equations become
\be
\label{eq:RG}
\dot J_1=J_1^2,~~~~\dot J_2=J_2^2 \, ,
\ee
 where  $J_1 ={1\over2} (J_{+} + \sqrt{J_{-}^{2} + \lambda_{F}^{2}})$; $J_2 ={1\over2} (J_{+} - \sqrt{J_{-}^{2} + \lambda_{F}^{2}})$; $J_{\pm}=J\pm \gamma$
and the ratio $\lambda_F/J_- =$ constant. Eq.~\ref{eq:RG} describes
the RG flow of two decoupled Kondo Hamiltonians with couplings $J_1$ and $J_2$.
Considering an initial condition with $\gamma=0$  and an antiferromagnetic Kondo coupling $J>0$, 
the equations render $J_1 > 0$ and $J_2 <  0$. Therefore, as temperature
is lowered, $J_1$ grows to the strong coupling regime while $J_2$ goes to zero, hence reducing to the standard one-channel Kondo problem. There is also another possible initial condition that has  $J < 0$ (ferromagnetic Kondo Hamiltonian). In this seemingly unfavorable case for the development of a Kondo regime, however, (\ref{eq:RG}) predicts the appearance of an antiferromagnetic coupling at low energies for $J_{1}$ while $J_{2}$ remains $< 0$, becoming eventually zero.
Note that in both cases, the final Kondo state is formed by the impurity coupled to
a combination of $m  = 0$ and $\pm 1$ modes from the bath.

\begin{figure}
\centerline{\includegraphics[width=3in]{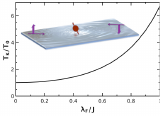}}
\caption{(Color online) Kondo temperature enhancement due to RSO grows exponentially with $\lambda_F$. Inset shows schematics of magnetic adatom on 2d electron gas system with RSO coupling where states for a given momentum have definite spin.}
\label{fig}
\end{figure}

Although the presence of RSO results on known physics in the Kondo regime, its effect on the Kondo temperature
is quite dramatic. Away from particle-hole symmetry conditions at the impurity, the DM term increases the Kondo 
coupling producing an exponential increase in the Kondo temperature given by:
\be
{T_K \over T_{0}} = \left(\frac{T_{0}}{D}\right)^{1-{J_{1}/J}} ,
\label{TK}
\ee
where $T_0$ is the Kondo temperature in the
absence of RSO and $D$ is the bandwidth cut-off \cite{Supplementary}. We should note that the change in the coupling $J$ produced
by RSO interactions is not compensated by changes in the effective density of states $\rho(\epsilon_{F})$ at the Fermi level. One can show that terms in the SWT that can renormalize $\rho(\epsilon_{F})$ are of two types: i) those that are independent of $\lambda_{R}$, 
and as such cannot eliminate the effect of the DM term in $J$; or ii) those that depend on $\lambda_{R}$ via the renormalized value of this
coupling. As this last renormalization is mainly due to the presence of the Hubbard term $U$, and the DM corrections depend 
on both $U$ and $\varepsilon_{d}$, the correction on $\rho(\epsilon_{F})$ can not generically
compensate the change in $J$. 
These effects can be substantial and significantly enhance $T_{K}$.  Figure \ref{fig} illustrates $T_K$ as function of the RSO parameter $\lambda_F$, changing in a clearly super-linear fashion, unlike the linear increase expected from density of state effects \cite{Malecki}.  For a Co atom adsorbed on graphene with SiO$_2$ as substrate \cite{Manoharan3}, for example, $T_0$ has been measured to be $ \approx 14$K; changing the substrate to Au/Ni has been shown to enhance the RSO strength to $\lambda_R \simeq 0.2$ eV \cite{Varykhalov}.  We estimate $\lambda_F/J \approx 0.3$, which would result in a 20\% increase for $T_K \approx 17$K.\@ 
We emphasize that this strong enhancement holds in the generic situation away from particle-hole symmetry at the impurity, suggesting that it should be   observed in quite general situations.

{\it Conclusion.}
In summary, we have analyzed the Kondo regime of an Anderson impurity model 
with RSO interactions in a 2d electron gas. Because of the broken  
SU(2) spin symmetry in the presence of RSO, the coupling between impurity and
band electrons occurs via a combination of different angular momentum modes in the bath.
The resulting two-band model produces a two-channel Kondo regime with ferro- and antiferromagnetic
coupling constants, having a standard one-channel Kondo physics as fixed point. We have shown, in
agreement with previous studies, that the effects of the RSO interactions vanish for an impurity in the particle-hole
symmetry point. Away from this point, however, RSO interactions introduce dramatic modifications
to the Kondo regime by generating a DM term that has been missed in previous studies. The DM term
is responsible for an exponential increase in the Kondo temperature. It is reasonable
to expect that these effects can be observed in experiments carried on with magnetic atoms placed on surfaces of
different metals \cite{Crommie1,Li,Hoch} or 2d semiconductors systems \cite{Goldhaber,Cronenwett} where the strength of the RSO coupling can be varied. 
Other interesting candidate systems are those with magnetic adatoms on graphene \cite{Manoharan3, Crommie6}, when supported by substrates that enhance the RSO interactions \cite{Dedkov,Varykhalov}, or on top of topological insulators \cite{Moore}. Work along these lines will be reported elsewhere.
 
{\it Acknowledgements}. We appreciate useful discussions with P. Brouwer, F. Guinea, K. Ingersent
and E. Kim. This work was supported by NSF SPIRE and MWN/CIAM grants and the AvH Stiftung. We acknowledge support from the Aspen
Center of Physics and the Dahlem Center for Complex Systems at Freie Universit\"at, Berlin, where parts
of this work were completed.

\end{document}